\documentclass[aps,prl,twocolumn,groupedaddress,showpacs]{revtex4}

\usepackage{graphicx}
\begin{document}

\title{Influence of the Dzyaloshinskii-Moriya exchange interaction on quantum phase interference of spins}

\author{W. Wernsdorfer$^1$, T.C. Stamatatos$^2$, and G. Christou$^2$}


\affiliation{
$^1$Institut N\'eel, CNRS and Universit\'e J. Fourier, BP 166, 38042 Grenoble Cedex 9, France\\
$^2$Department of Chemistry, University of Florida,
Gainesville, Florida 32611-7200, USA
}

\date{submitted 14$^{\rm th}$ July 2008}

\begin{abstract}
Magnetization measurements of a 
Mn$_{12}$mda wheel single-molecule magnet 
with a spin ground state of $S = 7$ show
resonant tunneling and quantum phase interference,
which are established by studying the tunnel rates as
a function of a transverse field applied along
the hard magnetization axis.
Dzyaloshinskii-Moriya (DM) exchange interaction
allows the tunneling between different spin multiplets.
It is shown that the quantum phase interference of
these transitions is strongly dependent on the direction 
of the DM vector.
\end{abstract}


\maketitle

Single-molecule magnets (SMMs), sometimes called molecular nanomagnets, 
consist of an inner magnetic core and 
a surrounding shell of organic ligands~\cite{Christou00} that can be tailored 
to bind onto surfaces or into junctions. SMMs come in a 
variety of shapes and sizes and permit selective 
substitution of the ligands in order to alter the 
coupling to the environment.  It is also possible to exchange 
the magnetic ions, thus changing the magnetic 
properties without modifying the structure and 
the coupling to the environment.  
SMMs combine the classic macroscale properties of a 
magnet with the quantum properties of a nanoscale entity. 
They display an impressive array of 
quantum effects that are observable up to higher and higher 
temperatures due to progress in molecular design, 
ranging from quantum tunnelling of magnetization~\cite{Friedman96,Thomas96,Sangregorio97} to 
Berry phase interference~\cite{Garg93,WW_Science99,WW_JAP02,WW_PRL05} and 
quantum coherence~\cite{Ardavan07,Bertaina08} 
with important consequences on the physics of spintronic devices~\cite{Bogani_NatMat08}. 

Up to now,  the spin system of an SMM has mainly been described by a 
single spin $S$, and the associated tunneling processes 
were transitions inside the multiplet of $S$~\cite{Christou00}. 
Recent studies in the field of molecular magnetism go beyond this giant-spin 
approximation, describing the molecule as a multi-spin
system~\cite{Carretta08}. In this case, the total spin $S$ of the molecule 
is not fixed, but several multiplets with different total 
spins appear and the number of allowed 
tunnel transitions and relaxation paths of the spin system 
increase considerably. For a simple multi-spin description
with symmetric exchange coupling between spins, 
tunneling between different multiplets is forbidden.
However, antisymmetric exchange coupling between spins, that is
the Dzyaloshinskii-Moriya (DM) interaction,  can lift the degeneracy
of energy level crossings belonging to different spin multiplets,
and tunneling and interference between these 
levels become allowed~\cite{Carretta08,Ramsey08,WW_condmat08,Bahr08}. 
DM interactions result in general from pairwise interactions 
of neighbouring spins, which do not have an inversion centre.
This condition is fullfilled  most of the time in SMMs even when
the entire molecule has an inversion centre~\cite{WW_condmat08,Elhajal05}. 

We present here resonant quantum tunneling measurements 
of a Mn$_{12}$mda wheel~\cite{Foguet_Albiol05}, which is a member of the
 [Mn$_{12}$(O$_2$CMe)$_{14}$(R-mda)$_8$] family of loops, 
 all with the same core and metal topologies but with R-substituted 
 mda$^{2-}$ groups~\cite{Rumberger05}. 
We show that this compound exhibits quantum phase interference
effects at all observed tunnel resonances and that the phase
of interference depends strongly on the direction of the DM vector $\vec{D}_{1,2}$.

The Mn$_{12}$mda (R = Me) wheel was prepared by the reaction of 
Mn(O$_2$CMe)$_2$.4H$_2$O and N-methyldiethanolamine (mdaH$_2$) in the 
presence of the organic base NEt$_3$, and crystallizes as dark red plate-like crystals in triclinic
space group $P\bar{1}$. 
Full details of the synthesis, crystal structure and magnetic characterization 
were presented elsewhere~\cite{Foguet_Albiol05}, establishing a
ground  state spin $S = 7$ that the wheel consists of alternating 
Mn$^{2+}$ and Mn$^{3+}$ ions, and therefore all subunits consisting of 
two neighboring Mn ions must lack an inversion centre, 
justifying therefore the possibility of DM interaction even though
the complete molecule has an inversion centre~\cite{WW_condmat08,Elhajal05}.
Although our Mn$_{12}$mda wheel is very similar to those in~\cite{Rumberger05,Ramsey08},
the hysteresis loops show resonant tunneling steps that
are much more narrow, allowing us to study in detail the quantum effects involved. 

The magnetization measurements were performed by using a 
micro-SQUID setup~\cite{WW_JAP95} on top of which a single crystal of 
Mn$_{12}$mda wheels was placed.
The field was aligned with the easy axis of magnetization using
the transverse field method~\cite{WW_PRB04}.

\begin{figure}
\includegraphics[width=.44\textwidth]{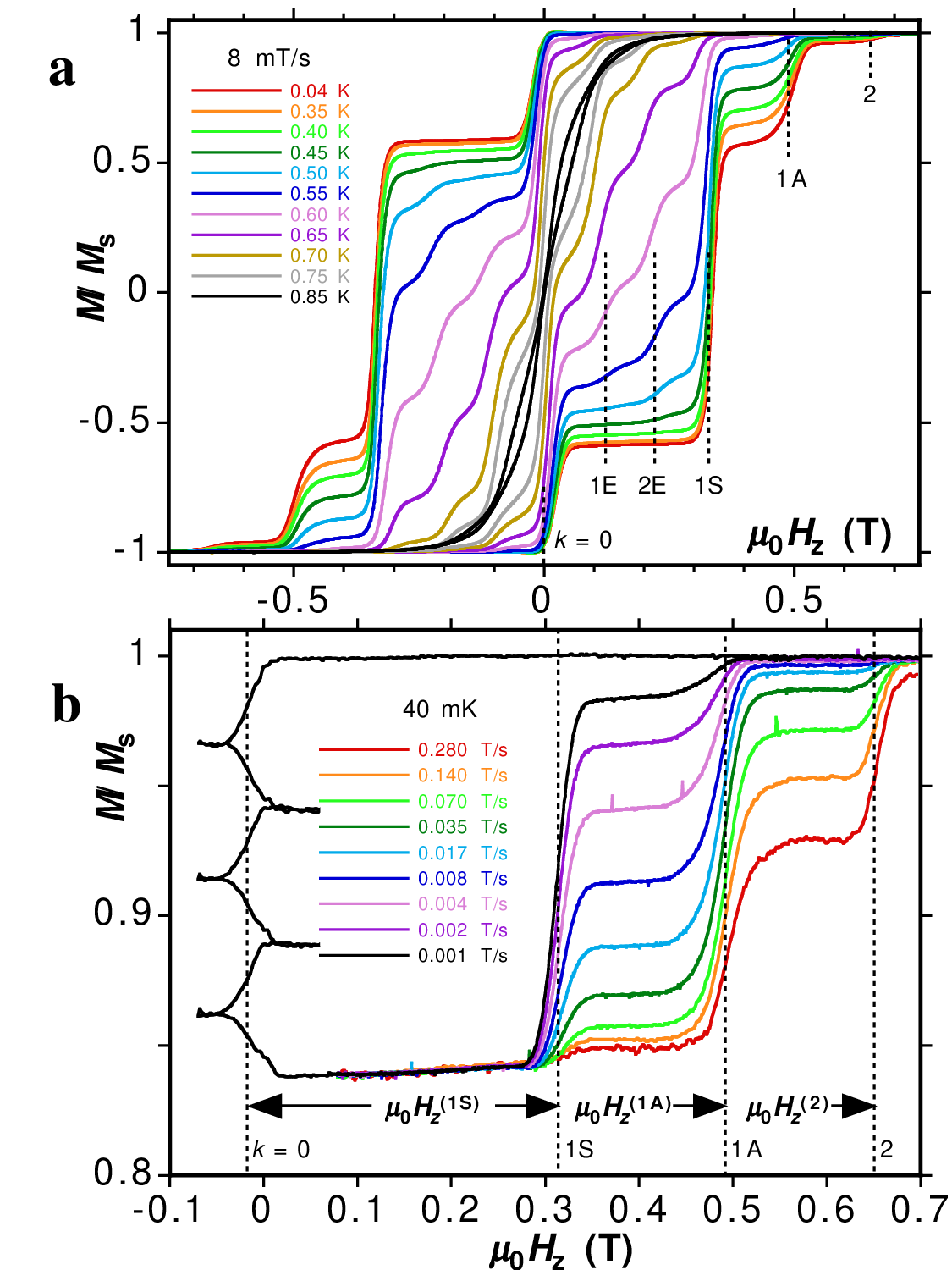}
\caption{(Color online) (a) Hysteresis loops of single crystals 
of Mn$_{12}$mda wheels 
at different temperatures and a constant field
sweep rate of 8 mT/s. 
(b) Minor hysteresis loops at 0.04 K.
The magnetization was first saturated at 1 T. 
After ramping the field to zero at 0.14 T/s, the field
was swept three times back and forth 
(between $\pm$0.07)
over the zero-field resonance $k=0$ with a sweep rate of 0.28 T/s.
Then, the field is quickly
swept back to 1 T at the indicated field sweep rates
leading to resonant tunneling at the transitions $k=$ 1S, 1A, and 2.
The corresponding field values are used to find the 
spin Hamiltonian parameters $D$ and $J$.}
\label{hyst_Mn12}
\end{figure}

Fig.~\ref{hyst_Mn12}a shows the temperature dependence of
the hysteresis loops of Mn$_{12}$mda wheels. The loops display a series of steps, 
separated by plateaus.  As the temperature is lowered, the
hysteresis increases because there is a decrease in 
the transition rate of thermally assisted tunneling~\cite{Friedman96,Thomas96}. 
The hysteresis loops become temperature independent below 0.3~K, demonstrating 
quantum tunneling at the lowest energy 
levels~\cite{Sangregorio97}.
Apart from the major steps, these hysteresis loops 
reveal fine structure in the thermally activated regime.
In order 
to determine precisely the resonance positions, we used the 
minor loop method described in~\cite{WW_PRB05c}. A typical example is
presented in Fig.~\ref{hyst_Mn12}b.

\begin{figure}
\begin{center}
\includegraphics[width=.44\textwidth]{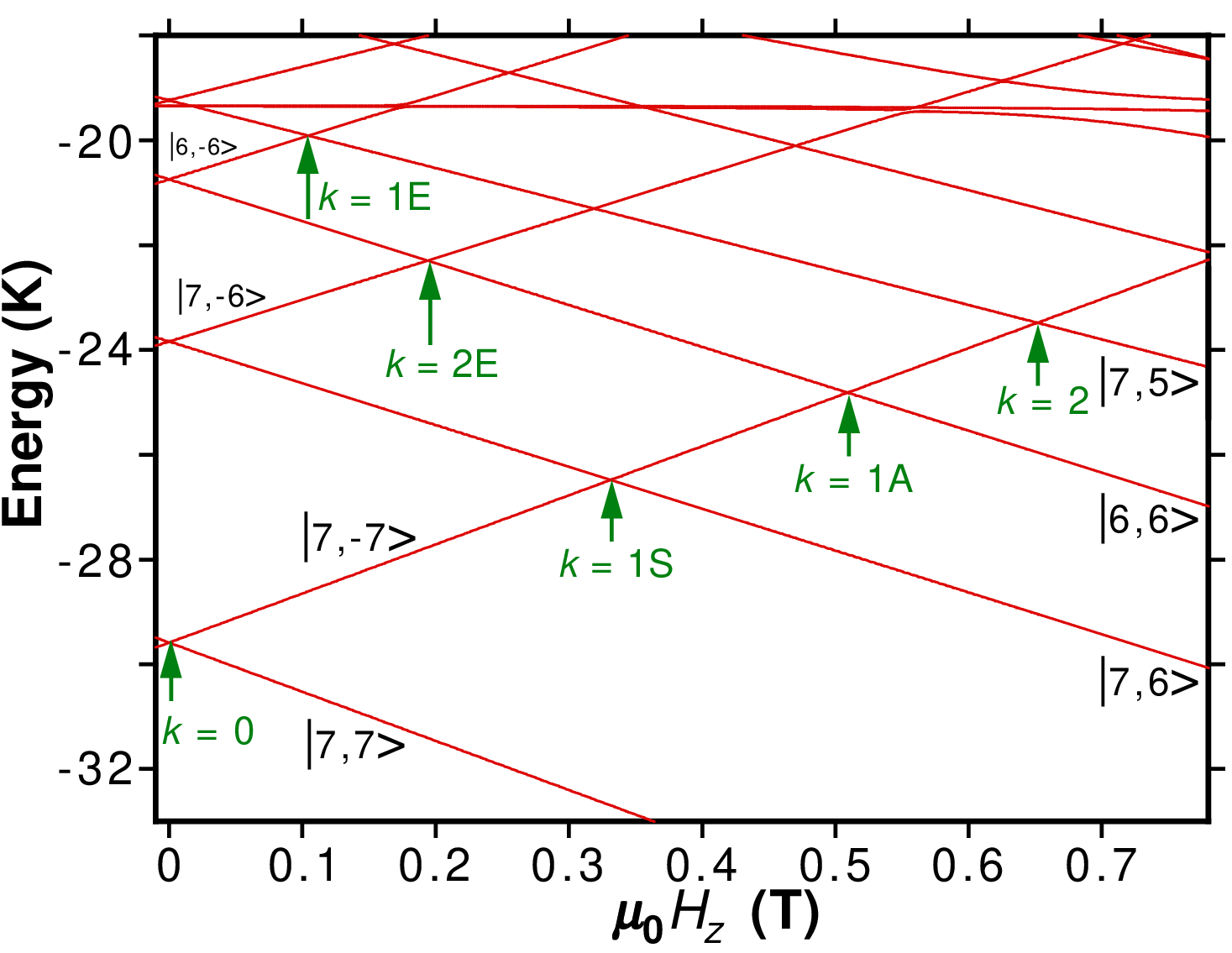}
\caption{(Color online) Zeeman diagram of the lowest energy levels 
used to explain the observed resonance tunnel transitions 
in Fig. 1. The field $H_z$ is along the easy axis of 
magnetization. 
The levels are labeled 
with quantum numbers $|S,M_S>$ and the observed level crossings are indicated with $k$.}
\label{Zeeman_Mn12}
\end{center}
\end{figure}

In order to explain the observed tunnel resonances and 
tunnel probabilities, and to study the influence of DM interaction 
on quantum phase interference, we model the 12-spin-system with
a simple dimer model of two ferromagnetically coupled spins  
$S_1=S_2=7/2$~\cite{Foguet_Albiol05,Ramsey08}. Although this
model can be questioned~\cite{WW_condmat08,Cano07}, 
it represents a useful simplification that keeps the required 
calculations manageable, has been found to describe well 
the lowest energy levels, and allows a qualitative discussion of the 
observed tunnel rates. The simple model employed does not 
affect the generality of the obtained conclusions 
about the influence of the DM interaction to be described.
Each spin $S_i$ is described by the spin Hamiltonian:
\begin{eqnarray}
    {\cal H}_i =  -D S_{i,z}^2 + E (S_{i,x}^2 - S_{i,y}^2) 
                          + \hat{\cal O}(4) - g \mu_{\rm B} \mu_0 \vec{S}_i\cdot\vec{H}
\end{eqnarray}
$S_{i,x}$, $S_{i,y}$ and $S_{i,z}$ are the vector components of the $i-$th spin operator 
and $g \approx 2$.
The first two terms describe the uniaxial anisotropy of the molecule, with longitudinal 
and transverse anisotropy parameters $D$ and $E$.  The third term contains higher order crystal field anisotropy terms.
The last term is the Zeeman interaction of the spin $\vec{S}_i$ with an external magnetic field $\vec{H}$.

The exchange interaction between the two spins can be described by
\begin{eqnarray}
        {\cal H}_{\rm ex} =  J \vec{S_1}\cdot\vec{S_2} + \vec{D}_{1,2} \cdot (\vec{S_1}\times\vec{S_2})
\end{eqnarray}
where the first term describes the isotropic Heisenberg exchange interaction with 
exchange constant $J$, and the second term is an antisymmetric DM interaction between the two spins
and the DM vector $\vec{D}_{1,2} $.

Exact diagonalization of the total spin Hamiltonian ${\cal H} = {\cal H}_1+ {\cal H}_2 + {\cal H}_{\rm ex}$ 
and use of $J =$ -0.435 K and $D =$ 0.985 K lead to the energy spectrum shown in Fig. 2.
The lowest lying spin states belong to the $S=7$ and $S=6$ multiplets.
Due to the ferromagnetic exchange, the first excited spin multiplet's $|S=6,M_S=\pm 6>$ doublet
is located at about 8.8~K above the ground state doublet $|S=7,M_S=\pm 7>$ in zero magnetic field.
Using $D=0.985$ K and $J=-0.435$ K, we can reproduce well the 6 observed
tunnel transitions. The resonances $k=0$, 1S, and 2 correspond to transitions between the
states of the $S=7$ multiplet, whereas $k=1A$, 1E, and 2E correspond to transitions 
between the $S=7$ and $S=6$ multiplets. The latter ones are not allowed unless
the antisymmetric DM interaction between the two spins is introduced (Eq. 2).

In order to get more insight into the tunnel process, we studied the tunnel resonances
as a function of a transverse field and used the Landau-Zener (LZ) 
method~\cite{WW_Science99,WW_JAP00_Fe8,WW_EPL00}.
We first placed a crystal of the Mn$_{12}$mda wheel in a high negative field $H_z$ to 
saturate the magnetization at 40 mK. We then swept the applied field at a constant 
rate $dH_z/dt$ over the $k=0$ resonance transition and measured the variation of 
magnetization using a micro-SQUID. For the other transitions, we used the 
minor loop method~\cite{WW_PRB05c} (Fig.~\ref{hyst_Mn12}b).
The fraction of molecules that 
reversed their spin was deduced from the step height, giving the LZ tunnel probability $P_k$
between two quantum states $m$ and $mÕ$. We deduced the 
corresponding tunnel splitting $\Delta_k$ using the LZ equation:
\begin{equation}
P_{k} = 1 - {\rm exp}\left\lbrack-
     \frac {\pi \Delta_{k}^2}
     {2 \hbar g \mu_{\rm B} |m - m'| \mu_0 dH_z/dt}\right\rbrack
\label{eq_LZ}
\end{equation}
Its validity can be tested by plotting $ \Delta_{k}$ as a function of $dH_z/dt$~\cite{WW_JAP00_Fe8,WW_EPL00}. 
We found that the LZ method is only applicable in the region of high sweep rates 
where $\Delta_{k}$ is independent of the field sweep rate. 
For the $k = 0$ resonance, this region is achieved for 
about  $\mu_0 dH_z/dt > 0.1$ T/s. 
The deviations from the LZ equation at lower sweep rates have been studied in 
detail~\cite{WW_PRL99,WW_EPL00} and are mainly due to reshuffling of internal fields~\cite{Liu01}. 
Note that $\Delta_{k}$ obtained at lower sweep rates 
always underestimates the real $\Delta_{k}$, it can therefore be used only as a lower-limit estimation~\cite{WW_condmat08}.

\begin{figure}
\begin{center}
\includegraphics[width=.44\textwidth]{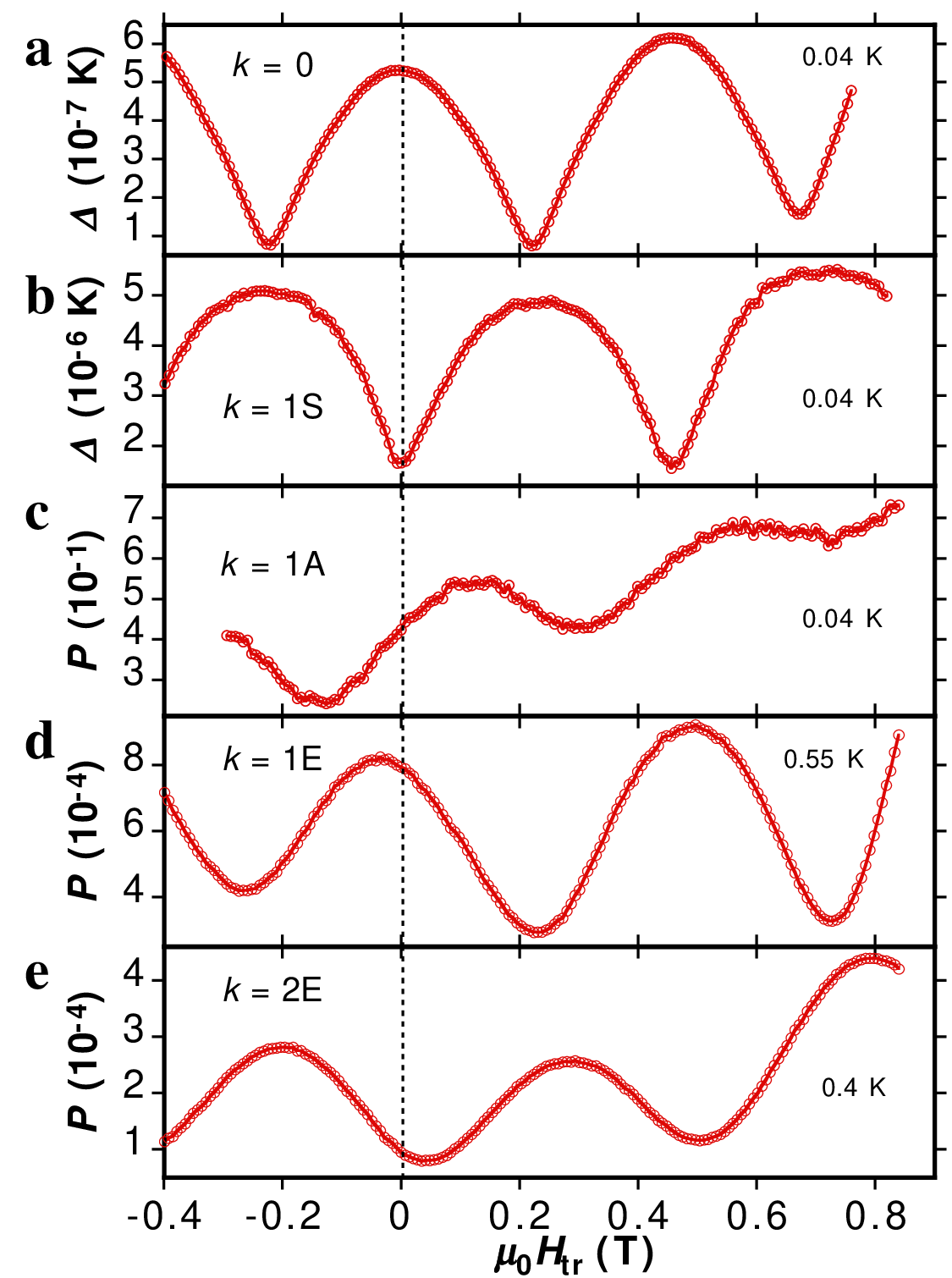}
\caption{(Color online)
Transverse field $H_{\rm tr}$ dependence of the tunnel splitting (a-b) and
the tunnel probability (c-e) for the indicated tunnel transitions.
$H_{\rm tr}$ was corrected by a mean internal transverse field of about 10 mT,
which was determined by measurements performed at positive 
and negative magnetization of the crystal.}
\label{delta_ex}
\end{center}
\end{figure}

Fig.  3 shows $\Delta_{k}$ as a function of a transverse field $H_{\rm tr}$, 
applied approximately along the hard axis of magnetization (x axis) 
and measured at $\mu_0 dH_z/dt =$ 0.56 T/s. The observed oscillations can be 
explained by quantum phase interference of two tunnel paths~\cite{Garg93}  
and has been observed in other SMMs~\cite{WW_Science99,WW_JAP02,WW_PRL05,Ramsey08,WW_condmat08}. 
We used the period of oscillation to determine the transverse anisotropy parameter $E=0.19$ K (Eq. 1).
At $\mu_0 dH_z/dt =$ 0.56 T/s, the LZ method is applicable only 
for $k =$ 0 and approximatively for $k =$ 1S. However, for $k =$ 1A the sweep rate was too slow
to apply the LZ method and we therefore plot only the tunnel probability in Fig. 3c.

Fig.  3d and 3e shows the tunnel probabilities for excited state tunnel transitions $k =$ 1E and 2E. 
Here, phonons first excite the spin from the ground state $|S,M_S> = (7,- 7>$
to the first or second excited spin states $|7,- 6>$ or $|6,- 6>$.
Then, during the LZ field sweep, the spin tunnels to $|6,6>$ or $|7,5>$, respectively.
Although this method can yield the activation energies and level lifetimes~\cite{WW_EPL00},
the tunnel splittings are difficult to deduce. Nevertheless, the tunnel probabilities 
$P_k$ can be found and studied as a function of transverse field (Fig. 3d-e), showing
clearly oscillations of $P_k$.

\begin{figure}
\begin{center}
\includegraphics[width=.44\textwidth]{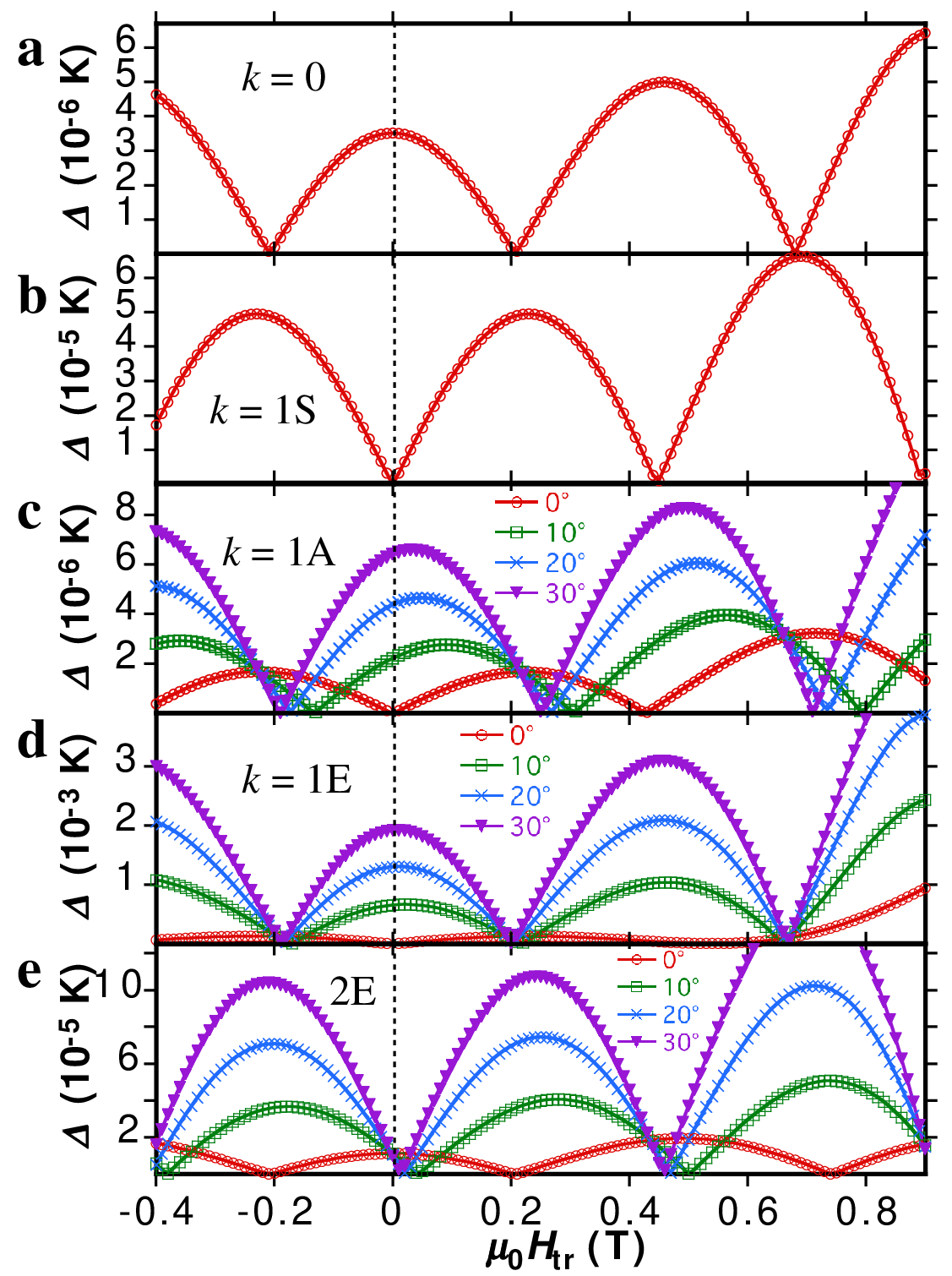}
\caption{(Color online) Calculated tunnel splitting 
for the indicated tunnel transitions $k$ as a function
of transverse field.
$\theta_{\rm DM}$ is indicated in (c$-$e) showing that the phases of
the oscillations depend strongly on $\theta_{\rm DM}$ .
The best agreement with the data in Fig. 3c-e is achieved for
$\theta_{\rm DM}$ = 10$^{\circ}$.}
\label{delta_theo}
\end{center}
\end{figure}

\begin{figure}
\begin{center}
\includegraphics[width=.44\textwidth]{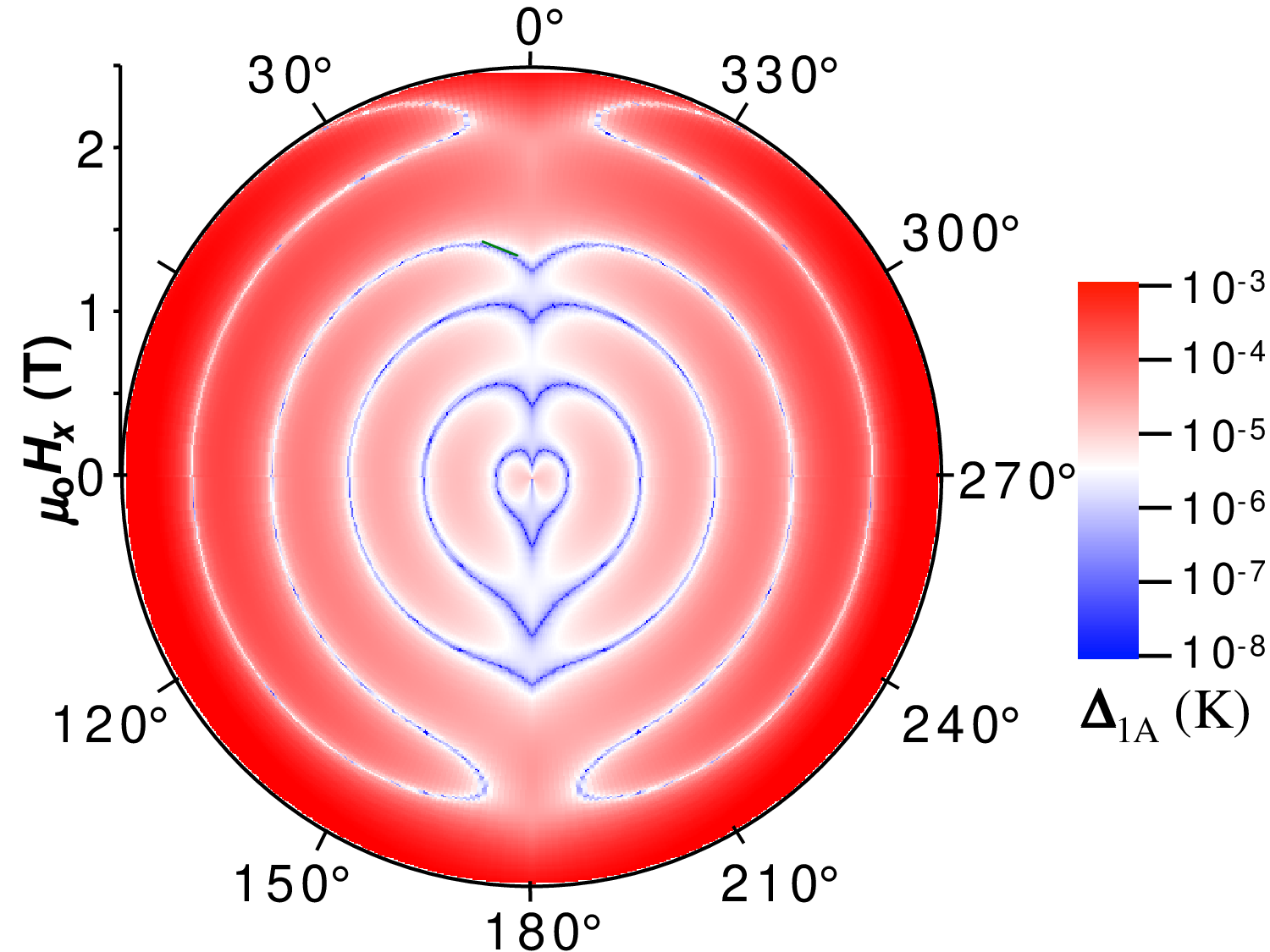}
\caption{(Color online) Color-scale representation of the calculated tunnel splitting 
for the tunnel transitions $k =$ 1A as a function of transverse field $H_z$ 
and the angle $\theta_{\rm DM}$ of the DM vector $\vec{D}_{1,2} $.}
\label{delta_angle}
\end{center}
\end{figure}

In multi-spin systems, transitions between different multiplets can be allowed by
DM interactions, that is, the observation of tunneling at $k=$ 1E, 2E, and 1A 
establishes the presence of a DM interaction in Mn$_{12}$mda wheels.
A very interesting observation is that the oscillations of the tunnel 
probabilities are not symmetrical with respect to the sign of the
transverse field ($P_k(H_{\rm tr} ) \neq P_k(-H_{\rm tr})$. This is in clear contrast to transitions between
states of the same multiplet ($k=$ 0 and 1S, see Fig. 3a-b). 
Numerical  diagonalization of the total spin Hamiltonian ${\cal H}$ shows
that the phase of the oscillation depends strongly on the orientation of
the DM vector  $\vec{D}_{1,2} $ (Eq. 2). Expressing $\vec{D}_{1,2} $
in terms of the modulus $|\vec{D}_{1,2} |$ and the usual polar angles
$\theta_{\rm DM}$ and $\varphi_{\rm DM}$ defined with respect to the $z$ axis, we found that 
(i) for small values of $|\vec{D}_{1,2} |$, $ \Delta_{k}$ does not depend on
DM interaction for a transition within a spin multiplet;
(ii) $\Delta_{k}$ depends strongly on
 $|\vec{D}_{1,2} |$ and $\theta_{\rm DM}$ for a transition between spin multiplets,
 whereas it hardly depends on $\varphi_{\rm DM}$. $\Delta_{k}$ is
 nearly proportional to $|\vec{D}_{1,2} |$ and the period of oscillation is
 close to those for transitions within a spin multiplet.
 Figs. 4 and 5 show a few examples of 
 $\Delta_{k}$ calculated with the Hamiltonian parameters given above, $|\vec{D}_{1,2} |$ = 0.03 K, 
 and several $\theta_{\rm DM}$ values.
 We find the best agreement for $\theta_{\rm DM} \approx 10^{\circ}$.
 
We would like to point out two deviations between the measurements
 and the dimer model, which we believe are due to the approximate nature of the latter.
 First, the experimental values of $\Delta_{k}$ for $k=$ 0 and 1S are about one order
 of magnitude smaller than the calculated ones. This discrepancy, also observed for a
 similar molecule~\cite{Ramsey08,WW_condmat08},  can be reduced
 by introducing $\hat{\cal O}(4)$ terms but it seems impossible to reproduce
 simultaneously the periods of oscillation and the values of $\Delta_{k}$.
 Second, the tunnel rates for $k=$ 1E, 2E, and 1A can be adjusted with the values of $|\vec{D}_{1,2} |$
 and the phase of oscillation with  $\theta_{\rm DM}$. However, we did not manage to find
 a  $\theta_{\rm DM}$ value that fits simultaneously the phases of all three transitions.
These deviations should motivate more theoretical work on the subject,
as well as extensions to more sophisticated models for the Mn$_{12}$ wheel 
involving two sets of six independent Mn spins.

In conclusion, we have shown for the first time how the DM interactions 
can affect the tunneling transitions and quantum phase interference
of a SMM. Of particular novelty and importance is the phase-shift 
observed in the tunnel probabilities of some transitions as a function 
of the DM vector orientation. Such observations are of importance to 
potential applications of SMMs that hope to take advantage of the 
tunneling processes that such molecules can undergo.  

We acknowledge discussion with B. Canals and S. Bahr.
This work was supported by the EC-TMR Network 
QuEMolNa (MRTN-CT-2003-504880), Magmanet, CNRS-ANR, 
Rh${\rm\hat{o}}$ne-Alpes funding,
and NSF.


\end{document}